\documentclass[a4paper]{aa}
\usepackage{graphicx}
\usepackage{txfonts}

\usepackage{url}
\usepackage{graphicx}
\usepackage[hyphenbreaks]{breakurl}
\usepackage{float}
\usepackage{subfig}
\usepackage{multirow}
\usepackage{amsmath}

\outer\def\gtae {$\buildrel {\lower3pt\hbox{$>$}} \over 
{\lower2pt\hbox{$\sim$}} $}
\outer\def\ltae {$\buildrel {\lower3pt\hbox{$<$}} \over 
{\lower2pt\hbox{$\sim$}} $}
\newcommand{\Msun}{$M_{\odot}$}

\newcommand{\kep}{\sl Kepler}

\newcommand{\tess}{\sl TESS}

\newcommand{\Plss}{P_{{\rm LS},s}}
\newcommand{\mPlss}{\langle P_{{\rm LS},s}\rangle}
\newcommand{\Dls}{\Delta_{\rm LS}}

\begin{document}

\title{Searching for stellar cycles on low mass stars using {\tess} data}
\author{Gavin Ramsay\inst{1}, Pasi Hakala\inst{2}, J. Gerry Doyle\inst{1}}

\authorrunning{Ramsay, Hakala \& Doyle}
\titlerunning{Stellar cycles on low mass stars}
\institute{Armagh Observatory and Planetarium, College Hill, Armagh, BT61 9DG, N. Ireland, UK\label{inst1}\and
Finnish Centre for Astronomy with ESO (FINCA), Quantum, Vesilinnantie 5, FI-20014, University of Turku, Finland\label{inst2}\\
\email{gavin.ramsay@armagh.ac.uk}}

\date{Accepted 9 July 2024}

\abstract {We have searched for stellar activity cycles in late low
  mass M dwarfs (M0--M6) located in the {\tess} north and south
  continuous viewing zones using data from sectors 1--61 (Cycle 1 to
  part way through Cycle 5). We utilise TESS-SPOC data which initially
  had a cadence of 30 min but reducing to 10 min in Cycles 3. In
  addition, we require each star to be observed in at least 6 sectors
  in each North/South Cycle: 1,950 low mass stars meet these
  criteria. Strong evidence was seen in 245 stars for a very stable
  photometric variation which we assume to be a signature of the stars
  rotation period. We did a similar study for Solar-like stars and
  found that 194 out of 1432 stars had a very stable modulation. We
  then searched for evidence of a variation in the rotational
  amplitude. We found 26 low mass stars showed evidence of variability
  in their photometric amplitude and only one Solar-like star. Some
  show a monotonic trend over 3--4 yrs whilst other show shorter term
  variations. We determine the predicted cycle durations of these
  stars using the relationship found by \citet{Irving2023} using an
  estimate of the stars Rossby number. Finally we find a marginally
  statistically significant correlation between the range in the
  rotational amplitude modulation and the rotation period.}


\maketitle

\section{Introduction}

Low mass stars account for nearly 3/4 of the stars in our local
neighbourhood \citep{Henry2006}. In spite of this, it is only in the
last few decades that a more intense focus has returned to these low
luminosity stars. Within the MV spectral type, stars have masses in
the range 0.08--0.6 \Msun\ (M9V--M0V), becoming fully convective around
early-to mid M.  The reason for the renewed interest in these stars is
due to exoplanets producing a greater transit depth in these small
stars making them easier to detect.  Furthermore, many of these stars
are active, especially fully convective stars
(e.g. \citet{Pettersen1989}), which provides insight into how fully
convective stars generate and sustain a global magnetic field, which
is of interest to the field of magnetism in general.

Moreover, the instruments to study stars of all kinds has radically
changed since the first exoplanets were detected. Prior to the launch
of {\sl Corot} and {\kep}, several groups attempted ground-based multi-telescope
observations of single stars (e.g. \citet{Rodono1986, Doyle1988a,
  Doyle1988b, Doyle1993}). What these series of space missions
facilitated (amongst many other things) was the measurement of the
rotation period for a large number of stars \citep{McQuillan2014} and
a comprehensive search for flares from stars with a wide range of
spectral type \citep{Davenport2016}. In turn, this allows for the
determination of how the stellar rotation period varies as a function
of age and mass \citep{Rebull2016} and how the rate of flares from an
active host star can affect the atmosphere of any orbiting exoplanet
\citep{Konings2022}.

Determining the rotation period of low mass stars is usually based
on the assumption that any modulation in the stars light curve is due
to the rotation of a starspot(s) which come into and out of view as the
star rotates. For stars whose rotation axis is viewed near face-on, it
is expected that starspot(s) would remain in view and show no
modulation unless the spot coverage varied over time. Initially it was
assumed that a single large spot group caused the modulation in a star's
flux, with any variations in the rotation flux profile due to changes
in the spot distribution or that additional starspots emerged.  However,
it has been shown that single filter photometry data can be reproduced
by a wide range of spot distribution, sizes, inclinations and spot
temperatures \citep{JacksonJeffries2013,Luger2021a}. On the
other hand, \citet{Luger2021b} argue that some degeneracies in the
spot distributions can be broken by analysing the light curve of many
stars which have broadly similar properties.

Although there is still some uncertainty in mapping the distribution
of spots over the surface of these stars, progress has been made in
searching for activity cycles on stars other than the Sun. The Solar
cycle lasts $\sim$11 yr (strictly speaking $\sim$22 yr) and manifests
itself through the changing number of sunspots; emission in the core
of the \ion{Ca}{\sc ii} H\&K lines; X-ray flux and the number of
flares and Coronal Mass Ejections. Given the timescales expected
(years), finding evidence for stellar activity cycles took time to
emerge. Systematic and regular spectroscopic observations of
Solar-like stars began during the 1960's with a focus on using the
\ion{Ca}{\sc ii} H\&K lines as an activity indicator
\citep{Wilson1968}. Subsequent studies using data such as these showed
that other stars exhibit activity cycles on a similar timescale to the
Sun (e.g. \citet{Baliunas1995} and references therein).  For lower
mass stars, an activity cycle of $\sim$7 yr was detected in the star
nearest to the Sun, Proxima Centauri (M5.5Ve), using optical, UV and
X-ray observations \citep{Wargelin2017}. Furthermore,
\citet{IbanezBustos2020} used Ca {\sc ii} H\&K lines to determine a
stellar cycle of $\sim$4 yr in the M4V Gl 729. \citet{Kuker2019}
reviewed observations which appear to show that fast rotating M dwarfs
have cycles of a year but increase to around four years for slower
rotators. However, fast rotators cannot be understood by an
advection-dominated dynamo model.

In principle, ground based optical photometric observations which are
stable and free from systematic trends can be used to detect activity
cycles as long as the timeline is sufficiently long. Since 2018 the
{\tess} satellite has gradually been building up a set of observations
of stars covering most of the sky. This is particulary true for stars
within $\sim11^{\circ}$ of the north and south ecliptic poles (the
`continuous viewing zones') since they can be observed for nearly one
year with a return visit one year later. However, since these light
curves are not flux calibrated, low amplitude, long term modulations
or trends can be difficult to detect in a robust manner.

An alternative means is to search for variations in the amplitude of
the rotational modulation over time \citep{Berdyugina2005}. Using this
approach, \citet{SuarezMascareno2016} determined that two K type stars
showed periods of 10.2 and 8.7 yr. {\citet{Reinhold2017} used {\kep}
  data to study the light curves of 23601 stars and used the
  Generalised Lomb Scargle Periodgram (LSP) to search for variations
  in the rotational amplitude with time and found amplitude
  periodicities in 3203 stars, with weak evidence that the cycle
  period increased for longer rotation periods. Most of the stars in
  that sample had spectral types earlier than MV stars. For a recent
  review of activity cycles on stars other than Sun see
  \citet{Jeffers2023}.

In this exploratory study we use {\tess} data of low mass stars
  located in the north and south continuous viewing zones on two
  occasions to assess how stable the dominant period in their light
  curves is. For stars which showed very stable periods we searched
  for amplitude variations which may reveal evidence of stellar
  activity cycles.

\section{The TESS sample}
\label{tess}

The {\sl TESS} satellite was launched into an Earth orbit with a
period of 13.7 d. It has four 10.5 cm telescopes that observe a
24$^{\circ}\times90^{\circ}$ instantaneous strip of sky (a sector) for
$\sim$27 d \citep[see][for details]{Ricker2015}.  In the first year
(Cycle 1) {\tess} covered most of the southern ecliptic hemisphere and
in the second year (Cycle 2) it covered the northern ecliptic
hemisphere, although with less sky coverage than Cycle 1 to avoid
contamination from stray Earth and Moon-light. In both Cycles, there
is an area around the ecliptic poles which are visible in each
sector. In Cycle 3 the field pointings mirrored Cycle 1 and in Cycle
4, some sectors observed in Cycle 2 were observed again in addition to
some fields not previous observed. In Cycles 1 and 2, full frame
images were obtained every 30 min, which increased to every 10 min in
Cycles 3 and 4. From the start, the Guest Observer programme was able
to populate targets which were observed in a 2-min cadence, with a 20
s cadence option becoming available in Cycles 3 and 4.

We used the {\it Gaia} EDR3 catalogue of stars within 100 pc
\citep{Smart2021} and using their $(BP-RP)$ colour and $MG$ absolute
magnitude we selected those which had an equivalent spectral type
later than $\sim$M0V and were on, or close to, the main sequence. We
then cross matched this sample with those stars which had {\tess}
light curves derived using the {\tess} Science Processing Operations
Center FFI Target List Products (SPOC, \citet{Caldwell2020}) which
uses the full-frame images to extract light curves (we searched light
curves from {\tess} sectors 0--61). Since our aim was to search for
variations in the amplitude of the main period, we then restricted our
sample to require a star to have been observed in at least 6
sectors. Finally, since the pixel size of {\tess} (21$^{``}$ per
pixel) is sufficiently large to include spatially nearby bright stars
to contaminate the resulting light curve, we filtered stars so the
Contamination Ratio \citep{Stassun2018} (i.e. the fraction of the flux
from other spatially nearby stars) is $<$0.1 . We show in Figure
\ref{gaia-hrd-all} the 1950 stars which are our M dwarf
sample. For a comparison, we also obtained a Solar-like sample using
the same criteria as for M dwarfs, where we took stars close to the
main sequence for $0.6<(BP-RP)<$1.1 (F5V--K2V,
\citet{PecautMamajek2013}): we identified 1432 stars and these are
shown as green dots in Figure \ref{gaia-hrd-all}.

\begin{figure}
    \centering
    \includegraphics[width = 0.45\textwidth]{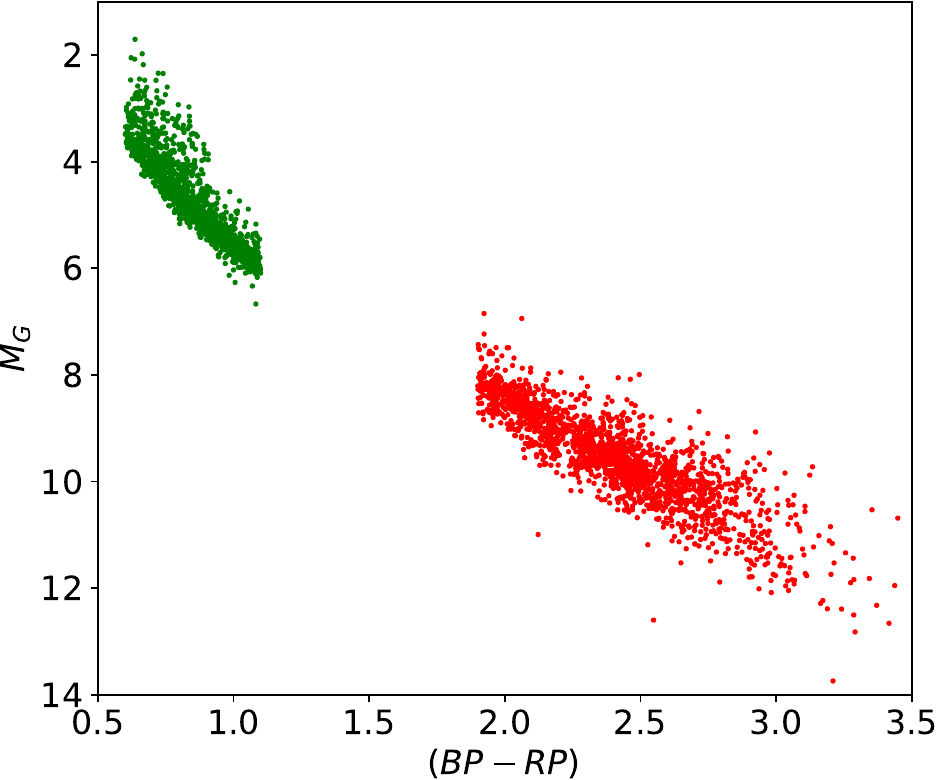}
    \caption{The {\it Gaia} HRD ($BP-RP$, $M_{G}$) of low mass stars (red)
      and Solar type stars (green) which were observed for at least
      six sectors in each {\tess} Cycle.}
    \label{gaia-hrd-all}
\end{figure}

\section{Data analysis}

Each sector of data was normalised so its mean was unity and then
clipped to remove data which was 5$\sigma$ deviation from the mean.
For each sector of data we used the LSP, as implemented in the {\tt
  VARTOOLS} suite of tools \citep{HartmannBakos2016,Zechmeister2009},
to search for evidence for periodic variability.  For each star we
determined the period of the highest peak in the LS power spectrum in
each sector of data in the range 0.08--10 d; its False Alarm
Probability (FAP) and the full amplitude (which we shorten to
amplitude for the rest of the paper) of the modulation for each sector
was determined.  We then obtained the median value for these
parameters for each star.  Since the lightcurves can still have
systematic trends present even after a global trend has been removed,
this can cause peaks in the LSP which are not astrophysical. In
addition, the observing window can cause peaks in the LSP, even at
periods which are a higher harmonic of the window length. Setting a
threshold for the FAP to determine if a star shows significant
variability is therefore not always clear cut
(e.g. \citet{Anthony2022}) and red-noise can result in peaks in the
LSP which technically have a FAP$<<$0.01
(e.g. \citet{Dorn-Wallenstein2019}).

\begin{figure*}
    \centering
    \includegraphics[width = 0.45\textwidth]{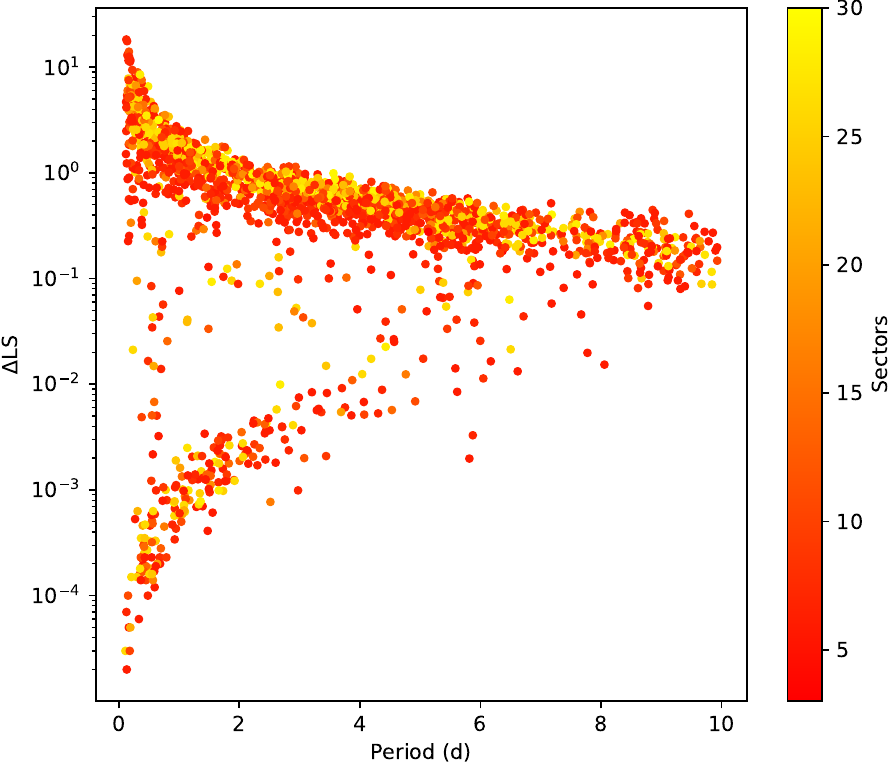}\hspace{5mm}
    \includegraphics[width = 0.45\textwidth]{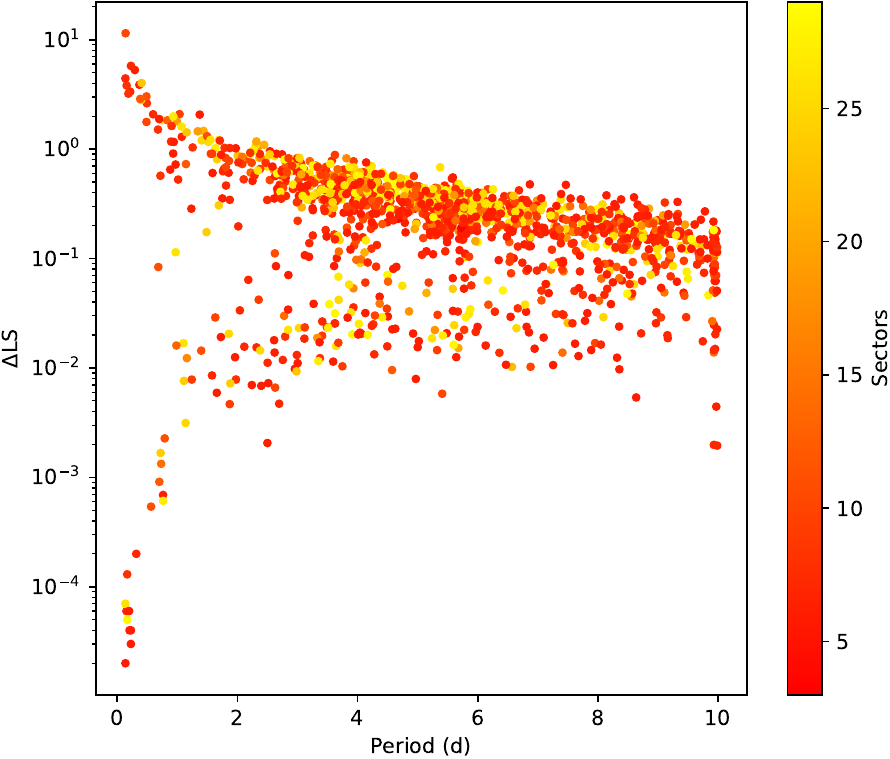}
\caption{Left hand panel: For all the 1950 M dwarf stars in the sample we
  determined the median period from their individual light curves and
  calculated $\Delta_{LS}$. Higher values of $\Delta_{LS}$ indicate a
  greater spread in the values of the period over the sectors in which
  they were determined. The colours of the points indicate how many
  sectors of data were analysed. Right hand panel: The same
  distribution but for the 1432 stars in our Solar-like sample.}
\label{gaia-hrd-delta}
\end{figure*}

We therefore use the approach of \citet{Lu2022},
who determined the spread in the period in different quarters of
{\kep} data using:

\begin{equation}
 \label{vartest}
  \Dls = \frac{1}{N_S \mPlss} \sum_{s}\left|\Plss
  -\mPlss\right|,
\end{equation}
\noindent
where $N_S$ is the number of sectors with measured $\Plss$, and
$\langle\cdots\rangle$ denotes the median value operator. We do not
necessarily expect the rotation period of active stars to be {\sl
  strictly} periodic, and therefore have very low values of $\Dls$,
since active regions can emerge and vanish at different latitudes over
the sequence of the observations, which together with differential
rotation of the star can lead to slightly different observed periods.

Since the LSP can identify half the real period when pulse profiles
are double peaked, we determined the median period and $\Dls$ and then
recomputed $\Dls$ so that if an individual sector was half the median
period (within a few percent) we used twice this value and
redetermined $\Dls$.  We show the median period and its spread
($\Dls$) for all 1950 stars in the left hand panel of Figure
\ref{gaia-hrd-delta}. There are two clear grouping: a band of points
starting at the lower left which have very low $\Dls$ values which
progress to higher values as the period increases. A second group of
points starts in the top left hand corner which have a large spread in
$\Dls$ which decreases as the period increases.

We examined the light curves and periodograms for stars which had
median periods shorter than 2 d but had values of $\Dls$ higher than
the strip of sources with low $\Dls$: we found evidence for some stars
appearing to be pulsators and others which had a longer period
modulation super-imposed on the shorter period seen in an earlier
Cycle of observations. This suggests that at least some stars above
the band of stars seen in the low $\Dls$ strip are probably real
variable stars but not isolated single M dwarfs which have
starspots. Of course many others are likely not real variable stars
and the LSP has just picked random peaks in the power spectrum which
are likely due to red noise.

We now make a comparison with the Solar-like sample discussed in Sect.
2. We applied exactly the same procedures as we did for the M dwarf
sample. Given Solar-type stars are expected to show (in general)
rotation periods much longer than a few days
(e.g. \citet{McQuillan2014}) we would not expect to see the same band
of points at the lower left -- unless their periods were spurious. We
show the distribution of the 1432 Solar-type stars in the right hand
panel of Figure \ref{gaia-hrd-delta}: we find the same spread of
points with high $\Dls$, but only a few stars in the lower left. There
are 17 Solar-like stars which appear to show very stable periods
(Figure \ref{gaia-hrd-delta}). We searched the {\tt
  SIMBAD}\footnote{\url{https://simbad.cds.unistra.fr/simbad/}}
database for information on these stars and found they were a mixture
of eclipsing binaries; eruptive variables and Red Clump low mass core
He burnging stars (e.g. \citet{Ruiz-Dern2018}). We conclude they are
not single star solar-like pulsators but other types of variable star,
which gives support to the view that the majority of stable periodic
low mass stars (left hand panel of Figure \ref{gaia-hrd-delta}) are
single low mass stars which have long lived starspots.

In selecting those sources with which to search for changes in
the amplitude of variability we observe that in both samples, there
is a `natural' split between most sources which have $\Dls$$>$0.05 and
$\Dls$$<$0.05. We therefore identified 245/1950 M dwarfs and 194/1432
solar-type stars which have $\Dls$$<$0.05 for the next stage of our analysis.

\begin{figure*}
    \centering
    \includegraphics[width = 0.95\textwidth]{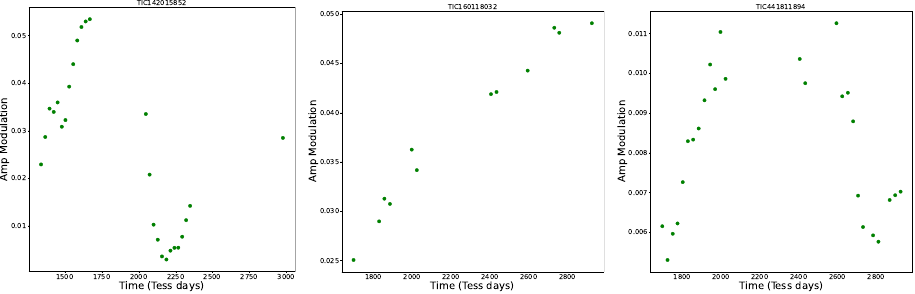}
    \caption{Three examples of low mass stars, TIC 142015852, TIC
      160118032, TIC 441811894, which have a variable rotational
      amplitude.}
    \label{longterm-amp}
\end{figure*}

\begin{figure*}
    \centering
    \includegraphics[width = 0.45\textwidth]{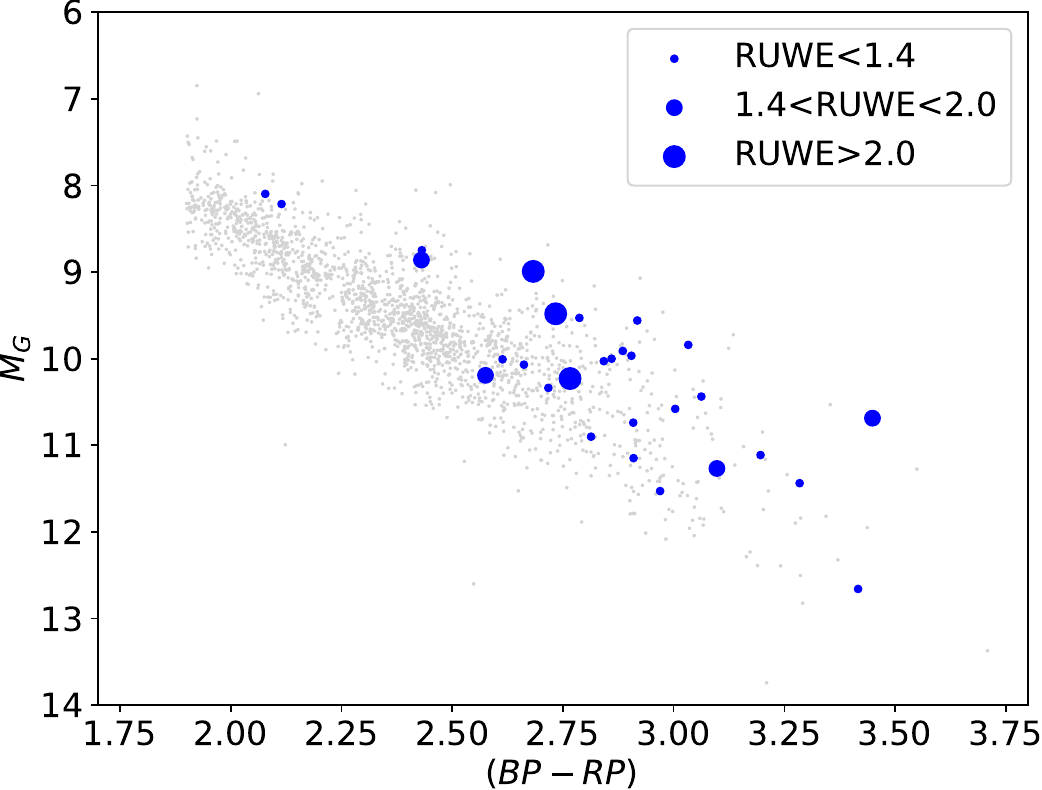}\hspace{5mm}
    \includegraphics[width = 0.45\textwidth]{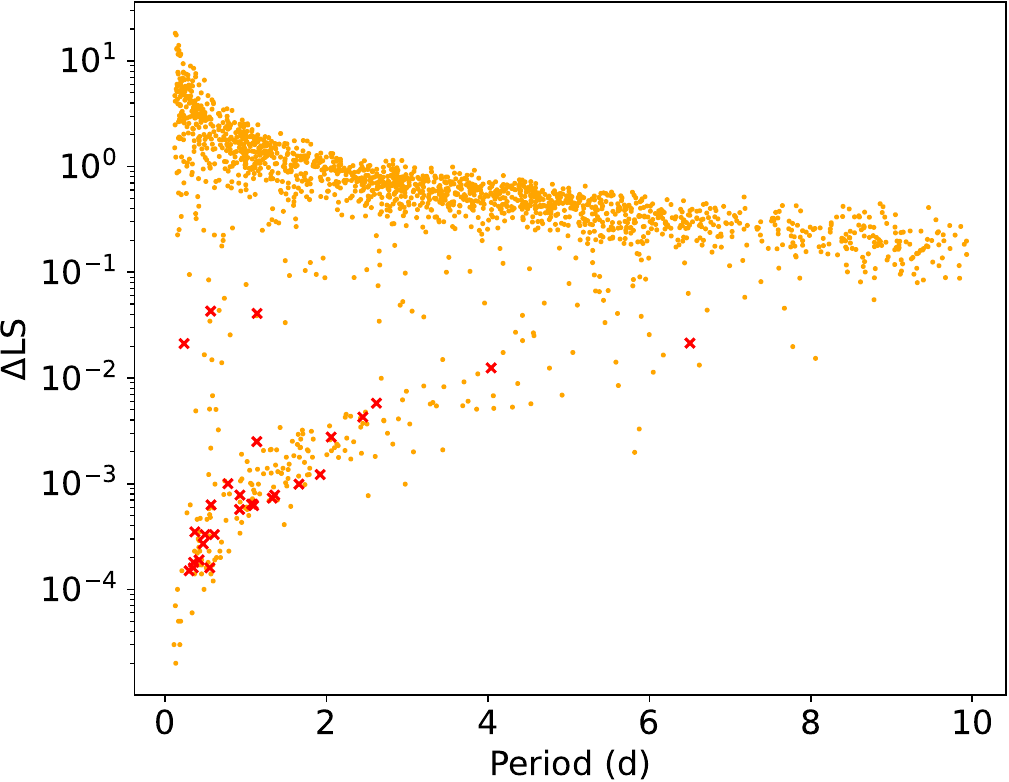}
    \caption{Left hand panel: The {\it Gaia} HRD ($BP-RP$, $M_{G}$) of low
      mass stars (grey) with those identified as showing significant
      variability of the rotational amplitude over time shown as
      circles whose size indicates their {\it Gaia} RUWE
      value (see text for details). Right hand panel: The low mass
      star sample (orange dots) with the 29 stars which were
      identified as having a rotational amplitude which varies
      significantly over time shown as a red $\times$ symbol.}
    \label{longVar-gaia-delLS}
\end{figure*}

\section{Searching for variations in the rotational amplitude}

Since the number of amplitude measurements per star is small (a
maximum of 30 spread over five years) we need to identify a suitable
statistical test which can robustly identify variations or trends in
the data.  In order to achieve this, we have used two different
non-parametric tests for variability in the time series i.e. the
Mann-Kendall ($MK$) test for monotonic trends in time series
\citep{Mann1945,Kendall1975} and a modified (non-parametric) version
of the $R$-statistic introduced in \citet{1993A&A...277..331B}.  The
$MK$ test is sensitive to monotonic (both linear and non-linear)
trends in the data, whilst the $R_{np}$-test measures the temporal
distribution of the signs of the residuals $r_i$ after subtracting the
mean level from the time series. We define $R_{np}$ as:
\begin{equation}
    R_{np} = {1 \over \sqrt{N-1}} \sum_{i=1}^{N-1}{r_i \over |r_i|}{r_{i+1} \over |r_{i+1}|}
\end{equation}
where $r_i$ are the time series residuals after subtracting the mean
level. $R_{np}$ follows a discrete normal distribution with a mean 0.0
and $\sigma=1.0$. Large positive values of $R_{np}$ correspond to
cases where a number of consecutive datapoints show the same sign
(either + or -) in their residuals after subtraction of the mean,
i.e. the constant value model does not fit the data.  The large
negative values of $R_{np}$ would imply systematically alternating
consecutive residuals with opposite signs and could only arise if the
data contained a short period that roughly matches the sampling
period. This case is not relevant here.  The probability distribution
for $R_{np}$ is discrete Gaussian, even if the data distribution would
not be known. This enables us to obtain reliable probabilities for
$R_{np}$ values, even in the case of a small number of points.

These two tests are sensitive to different types of variability
(monotonic trend vs.  correlated/systematic residuals for a constant
model), that can be present in the data. We therefore selected those
stars whose amplitude variation over time gave a FAP $<$0.0001 with
either the $MK$ stat {\sl or} the $R_{np}$-stat and had a minimum of 6
data points: this resulted in 29 low mass stars (if we choose FAP
$<$0.001 we obtain 53 stars). The details of these 29 stars are shown
in Table \ref{list-amp-variables} with three example light curves
showing the amplitude variation over time is shown in Figure
\ref{longterm-amp} (the amplitude variation of all 29 stars are shown
in Figure \ref{long-term-amp1}--\ref{long-term-amp3}). There is a
range of light curve characteristics such as TIC 142015852 which shows
a rise to maximum followed by a dip 1.5 yrs later; TIC 160118032 which
shows a continuous increase amplitude over 3 yrs and TIC 441811894
which shows a rise followed by a decline several years later. In
contrast, we found evidence for only one star in our solar-like sample
(TIC 38707949) which showed a general decrease in rotational amplitude
over a timescale of 2.5 yrs.

We also show the 29 stars in the {\it Gaia} HRD in the left hand panel of
Figure \ref{longVar-gaia-delLS} which shows that they are biased
towards the red or more luminuous parts of the lower main
sequence. This may imply the stars are younger compared to the overall
sample or they are members of binaries. To explore this further, we
indicate using the colour of the points the {\it Gaia} Renormalised Unit
Weight Error (RUWE) parameter for each of the 29 stars. This is a
measure of how much the photo-center of a star moves over the course
of the {\it Gaia} observations \citep{Lindegren2021a}. Initial indications
suggest that for stars with RUWE$>$1.4 the star is an unresolved
binary system \citep{Lindegren2021b}, although this does not preclude
that binaries can have RUWE$<$1.4 \citep{StassunTorres2021} and that
the distribution of RUWE can vary over the {\it Gaia} HRD
\citep{Penoyre2022}. We cross matched our 29 low mass stars with the
resulting catalogue of binaries within 100 pc from \citep{Penoyre2022}
finding that only two sources with the highest RUWE ($>$5.5) were in
their catalogue.  However, 7 out of these 29 stars have RUWE$>$1.4
suggesting they are possibly binary systems.

We folded and phase binned the data on the period identified for that
sector and show six examples in Figure \ref{fold}. Two of the stars
were in the binary sample derived by \cite{Penoyre2022}. Another one
star, TIC 38586438, is an eclipsing binary. We show the folded light
curves for three stars which are representative of the other stars in
the sample.

\begin{table*}
\caption{Details of the 29 low mass stars which show a significant variability in their rotational amplitude over time.}
  \resizebox{\textwidth}{!}{
    \begin{tabular}{rrrrrrrrrrrc}
\hline
  \multicolumn{1}{c}{TIC} &
  \multicolumn{1}{c}{RAJ2000} &
  \multicolumn{1}{c}{DEJ2000} &
  \multicolumn{1}{c}{Gmag} &
  \multicolumn{1}{c}{imag} &
  \multicolumn{1}{c}{MG} &
  \multicolumn{1}{c}{BP-RP} &
  \multicolumn{1}{c}{MedPer} (d) &
   \multicolumn{1}{c}{RUWE} &
   \multicolumn{1}{c}{$(V-K_{s})$} & 
   \multicolumn{1}{c}{$\tau$ (d)} &
   \multicolumn{1}{c}{$P_{predict,cyc}$ (yr)}\\
\hline
  33881250 & 66.342433 & -76.501968 & 14.23 & 13.35 & 10.4 & 3.06 & 1.141 & 1.33& 5.72 & 360 & 3.1\\
  38515801  &  62.551500 & -63.856128 & 14.57 & 13.68 & 10.0 & 2.86 & 1.072 & 1.19 & 5.50 & 370 & 3.7 \\
  38586438 & 64.090211 & -62.012890 & 12.83 & 12.03 & 8.9 & 2.43 & 0.556 & 1.53& 4.65 & 420 & 6.9\\
  142015852 & 96.591637 & -75.277925 & 12.45 &  & 8.1 & 2.08 & 2.622 & 1.24& 4.05 & 60 & 1.3\\
  142082942 & 98.259771 & -72.762015 & 14.23 & 13.34 & 10.9 & 2.81 & 0.924 & 1.10& 5.42 & 370 & 3.8\\
  149347629 & 83.648302 & -60.182846 & 14.13 & 13.25 & 9.8 & 3.03 & 0.566 & 1.10& 5.68 & 360 & 3.2 \\
  160118032 & 231.210928 & 83.991618 & 14.61 & 13.93 & 10.1 & 2.66 & 0.351 & 1.16& 5.03 & 400 & 5.2\\
  176980970 & 101.601200 & -70.634179 & 14.57 & 13.72 & 10.0 & 2.84 & 0.611 & 1.18& 5.34 & 380 & 4.1\\
  177313167 & 103.849793 & -74.048574 & 14.21 & 13.33 & 11.5 & 2.97 & 0.300 & 1.20& 5.53 & 370 & 3.6\\
  198557860 & 238.295269 & 62.595799 & 14.50 & 13.79 & 11.1 & 3.20 & 0.423 & 1.28& 5.69 & 360 & 3.2\\
  219773818 & 263.311581 & 63.207745 & 14.56 & 13.86 & 12.7 & 3.42 & 1.138 & 1.11& 6.26 & 350 & 2.2 \\
  230384382 & 285.730684 & 60.586065 & 14.72 & 14.04 & 10.2 & 2.77 & 0.497 & 5.56$^{*}$ & 4.93 & 400 & 5.5\\
  230395228 & 286.673233 & 63.754117 & 13.69 & 12.82 & 10.6 & 3.00 & 0.780 & 1.17 & 5.23 & 380 & 4.4\\
  233547261 & 281.738333 & 60.895959 & 14.46 & 13.77 & 10.0 & 2.61 & 2.451 & 1.14& 4.98 & 400 & 5.4\\
  271693722 & 109.549273 & -75.347401 & 14.71 & 13.86 & 10.3 & 2.72 & 1.923 & 1.19& 4.97 & 400 & 5.4\\
  278825715 & 100.316544 & -56.275061 & 13.46 & 12.56 & 9.5 & 2.79 & 6.503 & 1.24& & & \\
  294329266 & 109.049841 & -58.814441 & 13.50 & 12.64 & 9.5 & 2.73 & 0.370 & 2.44& 5.32 & 380 & 4.2\\
  300866996 & 117.229923 & -67.359987 & 14.32 & 13.37 & 10.7 & 3.45 & 0.355 & 1.84& 6.64 & 350 & 1.8\\
  306779230 & 121.428627 & -71.190802 & 14.41 & 13.52 & 9.6 & 2.92 & 1.329 & 1.16& 5.73 & 350 & 3.0\\
  339770560 & 111.121245 & -59.372994 & 13.86 & 12.99 & 9.9 & 2.89 & 1.098 & 1.15& & & \\
  341870349 & 276.732905 & 71.413962 & 13.22 & 12.61 & 8.7 & 2.43 & 4.041 & 1.27& 4.61 & 420 & 7.0\\
  349192028 & 108.836918 & -60.377753 & 12.86 & 12.14 & 8.2 & 2.11 & 2.059 & 1.30& 4.08 & 60 & 1.3 \\
  350559457 & 86.563846 & -55.797830 & 14.66 & 13.74 & 11.4 & 3.28 & 0.571 & 1.05& 6.16 & 350 & 2.4\\
  353898013 & 271.899854 & 56.324229 & 13.94 & 13.35 & 10.7 & 2.91 & 1.659& 1.19& 5.29 & 380 & 4.3\\
  375035131 & 99.647221 & -62.560617 & 14.77 & 13.85 & 10.0 & 2.90 & 0.928 & 1.09& 5.56 & 370 & 3.5\\
  382147080 & 80.523154 & -55.826228 & 13.52 & 12.65 & 9.0 & 2.68 & 0.473 & 6.24$^{*}$ & 4.95 & 400 & 5.5\\
  402897917 & 302.768145 & 68.500219 & 14.69 & 13.99 & 11.1 & 2.91 & 1.088 & 1.03& 5.33 & 380 & 4.2\\
  441734910 & 258.253191 & 73.934726 & 12.72 & 12.13 & 10.2 & 2.58 & 1.359 & 1.43 & 4.57 & 420 & 7.2\\
  441811894 & 266.958528 & 73.176166 & 14.34 & 13.40 & 11.3 & 3.10 & 0.236  & 1.72& 5.42 & 380 & 4.0\\
  \hline
\end{tabular}}
    \tablefoot{An asterix in
  the RUWE column implies it has been classed as a binary in
  \citet{Penoyre2022}, while the folded light curve of TIC 38586438
  shows it is an eclipsing binary (Figure \ref{fold}). We also show
  the $(V-K_{s}$) value from the TIC, the convective turnover time,
  $\tau$, and the predicted length of the activity cycle both derived
  from the relationships in \citet{Irving2023}. Stars TIC 33881250,
  149347629 and 441811894 are the stars with periods $<$2 d showing
  higher values of $\Dls$ than the other stars with crosses in the
  right hand panel of Figure \ref{longVar-gaia-delLS}.}
\label{list-amp-variables}
\end{table*}

In the right hand panel of Figure \ref{longVar-gaia-delLS} we show a
similar plot as the left hand panel of Figure \ref{gaia-hrd-delta} but
highlighting the location of the 29 low mass stars which we identified
as having variable amplitudes. Out of the 29 stars, 26 are located in
the band of stars with low $\Dls$ values. All the stars have periods
shorter than 7 days. There are three stars which have periods shorter
than 2 days and have $\Dls$ values much higher than the others. One of
these (TIC 33881250) has one sector where the period of the highest
peak differs from the others and has a much higher FAP, although
several flares could have interfered with the period
determination. Another two stars (TIC 149347629 and 441811894) show
two periods with the highest peak changing: this may indicate they are
pulsators. We therefore do not consider these stars in the next stage
of the analysis and are left with 26 stars.

\begin{figure*}
    \centering
    \includegraphics[width = 0.9\textwidth]{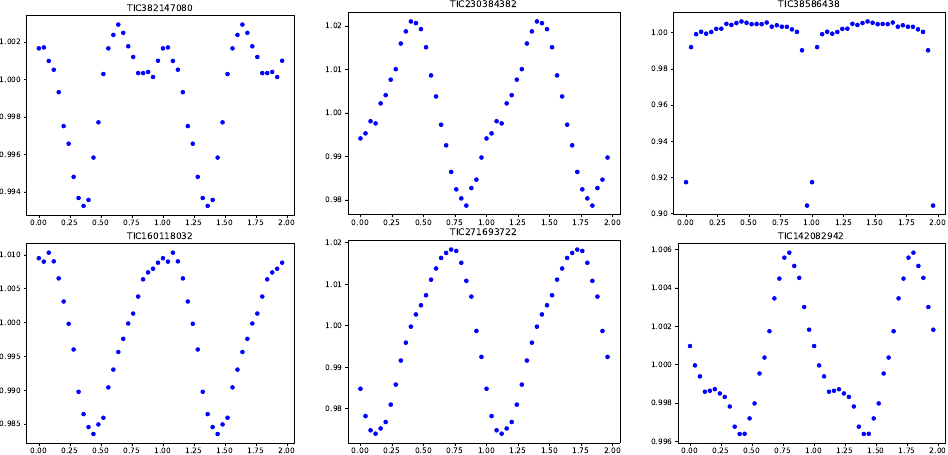}
    \caption{The first sector of data for six stars in our M dwarf
      sample which have been folded and binned on the period of the
      highest peak in the LSP -- the phase is arbitrary. The left hand
      and middle stars in the upper panel are in the binary sample of
      \citet{Penoyre2022} while TIC 38586438 appears to be an
      eclipsing binary. The shape of the folded light curves of TIC
      160118032, 271693722 and 142082942 are representative of other
      stars in the sample.}
    \label{fold}
\end{figure*}

\section{Expected length of stellar cycles in low mass stars}
\label{cyclelength}

Using measurements of the \ion{Ca}{\sc ii} H\&K lines using data taken
at Mt Wilson from the 1960's \citep{Wilson1968},
\citet{Brandenburg1998} found evidence that the ratio of the period of
the activity cycle, $P_{cyc}$, and the stellar rotation period,
$P_{rot}$, was related to the Rossby number ($R_{o}=P_{rot}/{\tau})$,
where $\tau$ is the convective turnover time.  More recently,
\citet{Irving2023} searched for a relationship between
$P_{cyc}/P_{rot}$ and $R_{o}$ for Solar-like and M dwarf stars. For
the latter they determined $P_{cyc}/P_{rot}$ = 3.6 R$_{o}^{-1.02}$
(see Fig 16 and eqn 4 of \citet{Irving2023}). Based on this
relationship we calculated what the predicted $P_{cyc}$ for the stars
in our sample would be. We assume that the period we measure from the
{\tess} data is the stars rotation period.

Because $R_{o}$ is not an observable quantity, we have to use a
proxy. One established means is that derived by \citet{Wright2018} who
use the $(V-K_{s})$ colour to derive $\tau$. However,
\citet{Irving2023} using work of \citet{Corsaro2021},
\citet{Landin2023} and \citet{PecautMamajek2013}, derived new
relationships for $(V-K_{s})$ and $\tau$. For $(V-K_{s})<4.6$ (stars
earlier than M3V) the resulting values of $\tau$ are similar to that
of \citet{Wright2018}. Following \citet{Irving2023} we use the
$\tau_{L}$ relationship for $(V-K_{s})<4.6$ and $\tau_{G}$ for
$(V-K_{s})>4.6$ (see Figure 13 of \citet{Irving2023}: for those
  stars near the boundary we chose the higher value). The values of
$\tau$ for each star are shown in Table
\ref{list-amp-variables}. Using $\tau$ and the period for each star we
compute $R_{o}$ and use the relationship of Irving et al. between
$P_{cyc}/P_{rot}$ and R$_{o}$ to determine the length of the activity
cycle of our stars. We show the predicted cycle lengths in Table
\ref{list-amp-variables} which range from 0.6--4.6 yrs: they are
longer by a factor of $\sim$2--7 compared to the values determined by
simply using the relationship of \citet{Wright2018}.

The amplitude variation over time is shown for all stars in Figures
\ref{long-term-amp1} and \ref{long-term-amp3}. In order to estimate
the length of the activity cycles from the amplitude time series data,
we experimented with fitting the amplitude time series with a Fourier
series models with different number of harmonic terms and also the
period as a free parameter. Such a problem is non-linear and thus we
carried out extensive MCMC fitting to derive limits/probability
distribution for any activity periods. However, we were not able to
obtain repeatable and reliable results.  This is due to the gaps in
the time series, and the length of the trends. However, it is clear
that some stars such as TIC 441734910 have a downward trend lasting
$\sim$3.5 yrs implying any activity cycles could be $\sim$8--10
yrs. The predicted activity cycle based on the relationship of
\citet{Irving2023} is 7.2 yrs. In contrast, TIC 142015852 shows an
amplitude variation which is consistent with a timescale of $\sim$3
yrs: the predicted cycle length is 1.3 yrs. This gives us modest
confidence that the variations which we observe are of the same order
as the predicted timescales for activity cycles.

\begin{figure}
    \centering
    \includegraphics[width = 0.45\textwidth]{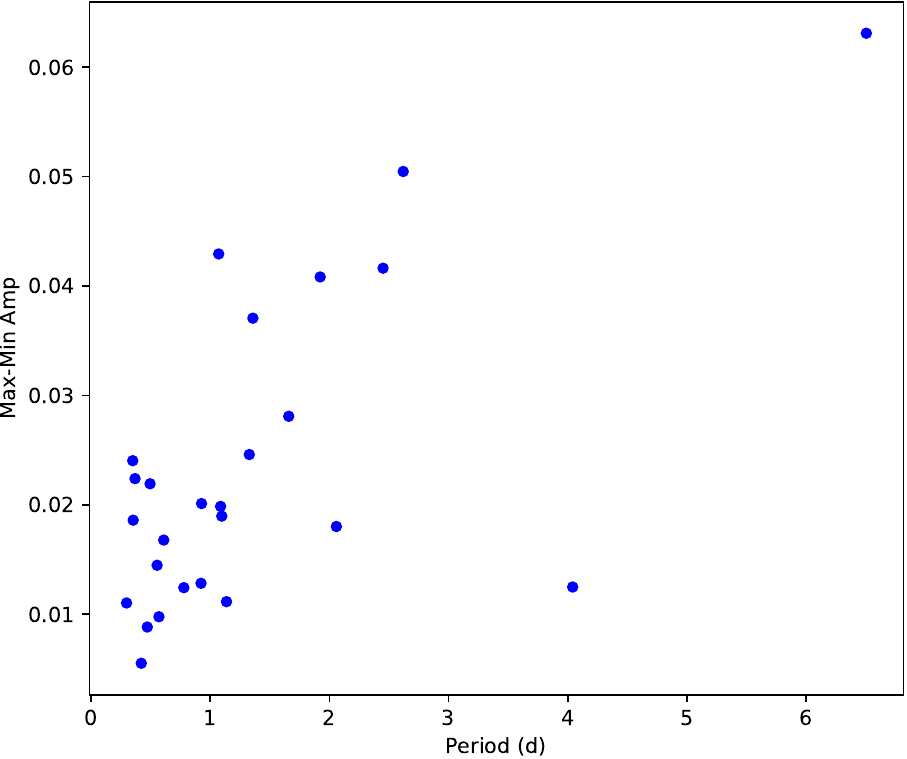}
    \caption{The variation in the amplitude of the modulation over the
      course of their observation as a function of modulation
      period. There is a marginally significant trend in the size of
      the amplitude variation over period.}
    \label{maxminperiod}
\end{figure}

\section{Amplitude variation as a function of rotation period}

We determined the range in the amplitude over time by simply taking
the difference between the maximum and minimum values for each of the
26 stars. The fractional mean amplitude `Max-Min' is 0.022, with a
range between 0.005 and 0.063.  We then searched for a correlation
between Max-Min and the stars mass, temperature (from the TIC,
\citet{StassunTorres2021}) and rotation period.  We used the Kendall
$\tau$ test to determine the significance of any correlation. We found
no evidence for a correlation between mass or temperature. However, we
found a FAP for the rank correlation between the amplitude variation
and modulation period of 0.00587 (Figure \ref{maxminperiod}). (A
  similar result was found when we compare the amplitude variation
  with Rossby number). Although this implies a 99.41 percent
probability that the correlation is real, it does not formally reach
the 3$\sigma$ level. The one outlying point is TIC 341870349
($P_{rot}\sim$4 d), although it is not obvious why this is the case.

Although statistical confidence in the correlation does not formally
reach the 3$\sigma$ level, we note that stars showing a larger range
in the amplitude of modulation are likely to have a greater asymmetry
in the distribution of spots than stars showing a smaller
modulation. It may imply stars which are faster rotating (and
therefore likely young) have spots more uniformly distributed and
covering a larger fraction of their photosphere.

\section{Discussion}

There is now robust evidence from spectroscopic observations that
Solar-like stars can show stellar activity cycles similar in duration
to that found in the Sun. This is because of the length of
observations, such as initiated at Mt Wilson \citep{Wilson1968}, but
also more recent high precision spectroscopy from instruments such as
HARPS \citep{Lovis2011}. Obtaining similarly robust claims for other
stars such as M dwarfs using photometric measurements is difficult
since all-sky precise photometry is still challenging
(e.g. \citet{Stubbs2006}). We also note that most claims of
significant detections of periods in photometric data are made using
the LSP which assume that data have a Gaussian noise profile whereas
almost all such data streams contain red noise which result in FAP's
which are far lower than they are in reality
(e.g. \citet{Littlefair2017}). Similarly, \citet{BasriShah2020}, using
simulations of starspot distributions, show that `short' activity
cycles (of the order of several tens or more rotation cycles) are
simply due to random processes.

In this exploratory study, we have gone to great lengths to identify
stars which have very stable photometric periods over a timescale of
several years. As a result we have found 26 low mass M dwarfs which
show evidence for a significant variation in the amplitude of their
rotational modulation over a time interval of nearly five years. There
are some stars whose amplitude variation points to a long term cycle
of up to ten years and there are other stars whose timescale is
$\sim$3--4 yrs. Given the relatively short rotation periods, we do not
consider these to be equivalent of spurious activity cycles noted by
\citet{BasriShah2020}.

Compared to Solar-like stars, the number of low mass stars (MV and
later) which have robustly determined activity cycles from spectra is
much fewer, partly because they are fainter in the blue where the
activity marker \ion{Ca}{\sc ii} H\&K lines are
present. \citet{SuarezMascareno2016} used photometric data obtained
using the All Sky Automated Survey (ASAS) to search for activity
cycles and found 25 early M and 9 mid M spectral types -- these
samples were biased towards relatively long rotational periods, a
median of 33.4 d (early M) and 86.2 d (mid M). In the sample of
\citet{Irving2023} which was mentioned in Sect. \ref{cyclelength}, of
their 15 stars, four had periods shorter than 6 d and only one with a
period shorter than 2 d. Our sample of 26 stars therefore reveals a
set of low mass stars which are rapidly rotating and have potential
activity cycles.

\section{Conclusion}

In this pilot study, we examined the {\tess} light curves of 1950 low
mass and 1432 Solar-type stars. We found that only 245 (12.6\%) low
mass stars showed stable periods over many sectors compared to 194
(13.5\%) Solar-type stars -- we assume that these periods are a
signature of the stars rotation period. However, there are many more
low mass stars which show very stable periods at periods shorter than
$\sim$4 d than Solar-type stars. This implies that the low spread in
measured periods of the short period M dwarfs is real and not produced
by sampling or any artefacts in the data. We note that the FAP of
periods detected using the LSP can give periods which may appear to be
significant at the 1\% level (which is commonly used in the
literature) which are in fact due to red noise.

We searched for variations in the amplitude of the rotation period
using two statistical tests and find that 29 low mass stars and one
Solar-type star shows evidence for significant variability. The
{\tess} light curves of stars in the continuous viewing zones all have
gaps of one year. This makes determining any meaningful estimate of
the period of variation difficult. However, they are not widely
inconsistent with the predicted cycle length which is based on the
stars Rossby number.

One mission which will observe the same field continuously is ESA's
{\sl PLATO} misssion which is due to be launched in Dec 2026 and will
make an initial pointing in the southern hemisphere lasting at least 2
yrs \citep{Nascimbeni2022} covering a continuous area of 1037
deg$^{2}$. Since this field will have been observed a
number of times using {\tess} it will provide the means to study the
modulation amplitude of thousands of stars over a time interval
comparable or longer to the expected activity cycles. 

Another means to fill the gaps is ground based wide-field optical
surveys such as {\sl GOTO} \citep{Steeghs2022,Dyer2022}, or {\sl
  ATLAS} \citep{Tonry2018}, which have telescopes in both hemispheres
are perfectly placed to fill in the gaps when {\tess} is not able to
observe parts of the sky, if they make extended observations of the
same fields every few months. It would also be interesting to
  determine if there is any variation in the X-ray flux of the 26
  stars found to show long term variability in their rotational
  amplitude. We hope that eRosita \citep{Predehl2021} can resume
  operations in the near future.
\vspace*{5mm}

\begin{acknowledgements}

This paper includes data collected by the TESS mission, for which
funding is provided by the NASA Explorer Program. This work also
presents results from the European Space Agency (ESA) space mission
{\it Gaia}. {\it Gaia} data is being processed by the {\it Gaia} Data Processing and
Analysis Consortium (DPAC).  Funding for the DPAC is provided by
national institutions, in particular the institutions participating in
the {\it Gaia} MultiLateral Agreement (MLA). The {\it Gaia} mission website is
https://www.cosmos.esa.int/gaia.  The {\it Gaia} archive website is
https://archives.esac.esa.int/gaia. Armagh Observatory \& Planetarium
is core funded by the N. Ireland Executive through the Dept. for
Communities. JGD would like to thank the Leverhulme Trust for a
Emeritus Fellowship.\\

\end{acknowledgements}

\begin{appendix}

\section{Long term variability of amplitude}

\begin{figure*}
    \centering
    \includegraphics[width = 0.9\textwidth]{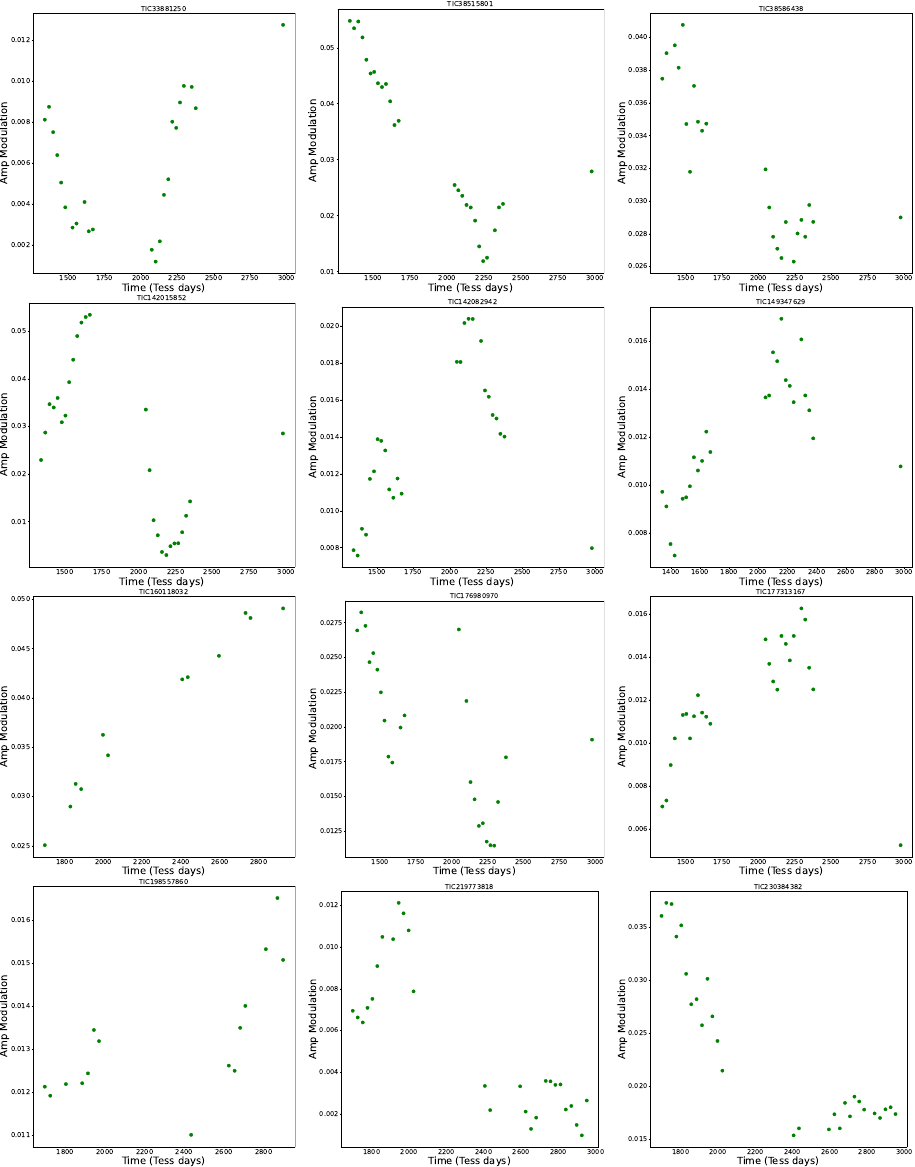}
\caption{The amplitude of the rotational modulation of low mass stars which we have identified as showing significant variability.}
\label{long-term-amp1}
\end{figure*}

\begin{figure*}
    \centering
    \includegraphics[width = 0.9\textwidth]{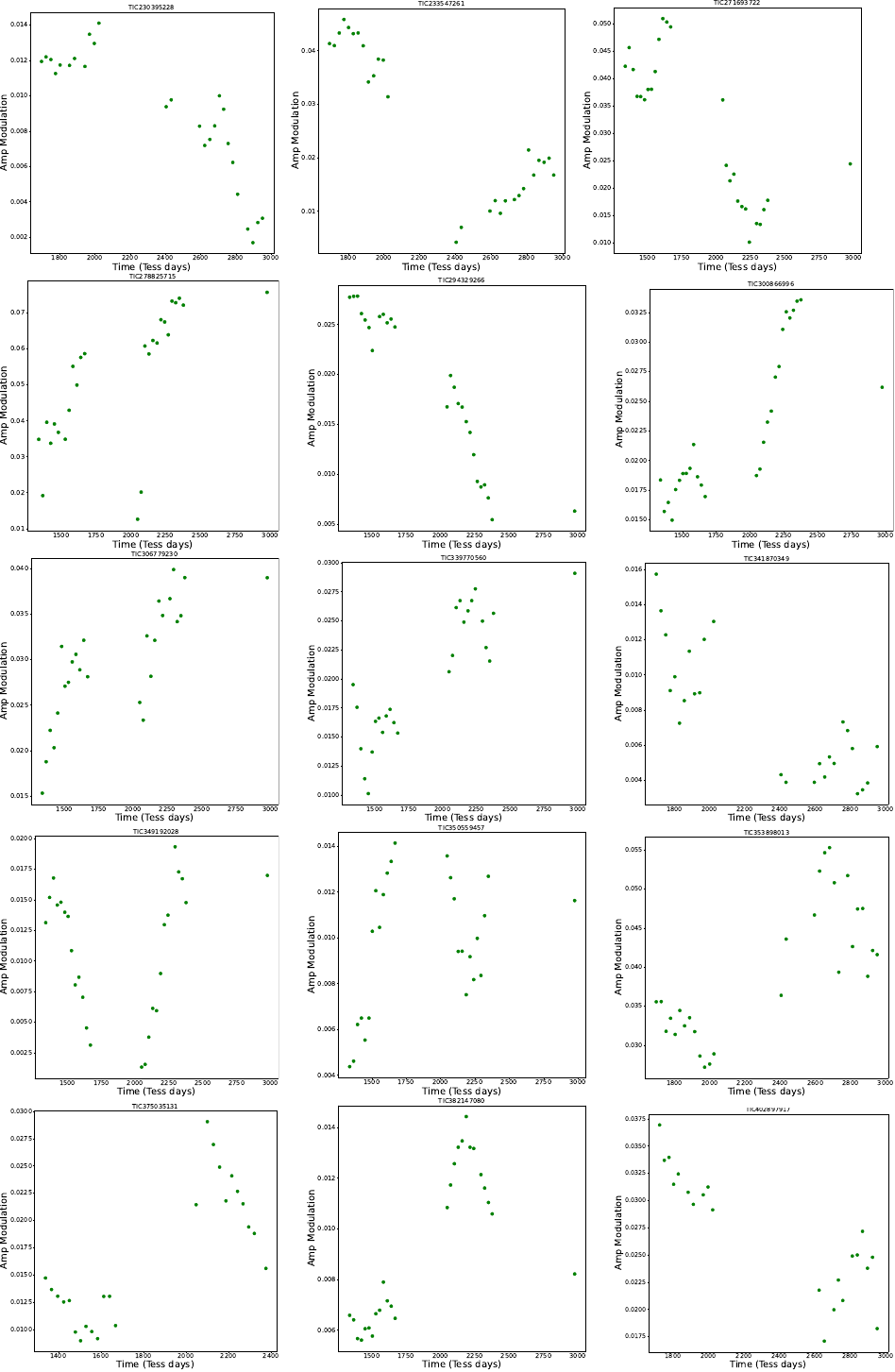}
\caption{The amplitude of the rotational modulation of low mass stars which we have identified as showing significant variability (Cont).}
\label{long-term-amp2}
\end{figure*}

\begin{figure*}
    \centering
    \includegraphics[width = 0.65\textwidth]{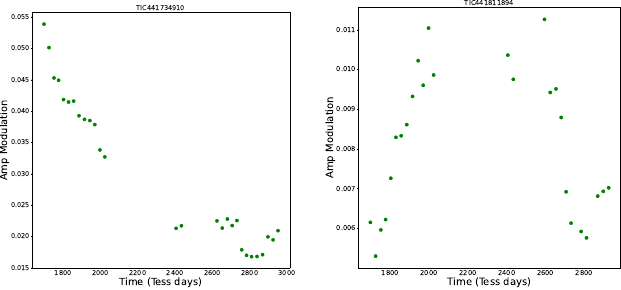}
\caption{The amplitude of the rotational modulation of low mass stars which we have identified as showing significant variability (Cont).}
\label{long-term-amp3}
\end{figure*}

\end{appendix}

\end{document}